\documentclass[numbers,final]{aipproc}
\layoutstyle{8x11double}
\usepackage{graphicx}
\usepackage{amsmath}
\usepackage{amsfonts}
\usepackage{amssymb}
\usepackage{feynmf}
\usepackage{xspace}

\allowdisplaybreaks

\def\siml{{\ \lower-1.2pt\vbox{\hbox{\rlap{$<$}\lower6pt\vbox{\hbox{$\sim$}}}}\ }}
\def\simg{{\ \lower-1.2pt\vbox{\hbox{\rlap{$>$}\lower6pt\vbox{\hbox{$\sim$}}}}\ }}

\def \m2   {\mu^{2 \epsilon}}

\newcommand{\ORd}[1]{{\mathcal O}\Bigl(#1\Bigr)}

\newcommand{\Mae}[3]{\bigl\langle#1\bigr\rvert#2\bigr\rvert#3\bigr\rangle}

\newcommand{\df}{\mathrm{d}}
\newcommand{\img}{\mathrm{i}}

\newcommand{\tr}{\mathrm{tr}}

\newcommand{\de}{\delta}

\newcommand{\w}{\omega}

\newcommand{\cG}{{\mathcal G}}
\newcommand{\cJ}{{\mathcal J}}

\newcommand{\cP}{{\mathcal P}}

\newcommand{\bn}{\bar{n}}
\newcommand{\bnP}{\overline {\mathcal P}}

\newcommand{\bnslash}{\bar{n}\!\!\!\slash}

\newcommand{\nn}{\nonumber}

\newcommand{\lqcd}{\Lambda_\mathrm{QCD}}

\newcommand{\SCETa}{\ensuremath{{\rm SCET}_{\rm I}}\xspace}
\newcommand{\SCETb}{\ensuremath{{\rm SCET}_{\rm II}}\xspace}

\newcommand{\beq}{\begin{equation}}
\newcommand{\eeq}{\end{equation}}
\newcommand{\beqa}{\begin{eqnarray}}
\newcommand{\eeqa}{\end{eqnarray}}

\def\dblone{\hbox{$1\hskip -1.2pt\vrule depth 0pt height 1.6ex width 0.7pt
\vrule depth 0pt height 0.3pt width 0.12em$}}

\begin{document}
\title{Fragmentation inside an identified jet}
%\preprint{TUM-EFT 2/09}

\author{Massimiliano Procura}{
address={Physik-Department, Technische Universit\"at M\"unchen, D-85748 Garching, Germany},
altaddress={Institute for Theoretical Physics, University of Bern, CH-3012 Bern, Switzerland}}

\author{Iain W. Stewart}{
address={Center for Theoretical Physics, Massachusetts Institute of Technology, Cambridge MA 02139, USA}}

\begin{abstract}
Using Soft-Collinear Effective Theory (SCET) we derive factorization formulae for semi-inclusive processes where a light hadron $h$ fragments from a jet whose invariant mass is measured. Our analysis yields a novel ``fragmenting jet function'' ${\cG}_i^h(s,z)$ that depends on
the jet invariant mass $\sqrt{s}$, and on the fraction $z$ of the large light-cone momentum components of the hadron and the parent parton $i$.
We show that ${\cal G}^h_i(s,z)$ can be computed in terms of perturbatively
calculable coefficients, ${\cal J}_{ij}(s,z/x)$, integrated against standard
non-perturbative fragmentation functions, $D_j^{h}(x)$. 
%We also show that
%$\sum_h \int_0^1 dz\: z\,{\cal G}_i^h(s,z)$ is given by the inclusive jet
%function $J_i(s)$ which is perturbatively calculable in QCD.
Our analysis yields a simple
replacement rule that allows any factorization theorem depending on a
jet function $J_i$ to be converted to a semi-inclusive process with a
fragmenting hadron $h$.
\end{abstract}

\date{\today}

\maketitle

%%%%%%%%%%%%%%%%%%%%%%%%
\section{Introduction}

In single-inclusive hadron production, an
energetic parton $i=\{u,d,g,\bar u,\ldots\}$ turns into an observed energetic hadron $h$ and
accompanying hadrons $X$.
The fragmentation function $D_i^h(z)$ characterizes the factorization theorems at leading power~\cite{Collins:1989gx} for high-energy semi-inclusive processes where no properties of the accompanying hadrons are probed. $D_i^h(z)$  encodes the non-perturbative information on how the parton $i$ (either a gluon or a flavor of quark or antiquark) produces the hadron $h$ which carries a fraction $z$ of the initial parton large light-cone momentum component~\cite{Collins:1981uw, Collins:1992kk, Mueller:1978xu, Georgi:1977mg, Ellis:1978ty,
  Curci:1980uw}.

Defining $n^\mu=(1,0,0,1)$ and $\bn^\mu=(1,0,0,-1)$, the light-cone components
of a generic four-vector $a^\mu$ are denoted by $a^+=n \cdot a$ and $a^-=
\bar{n} \cdot a$ where $n^2=\bar{n}^2=0$ and $n \cdot \bar{n}=2$. With $a_\perp^\mu$
we indicate the components of $a^\mu$ orthogonal to the plane spanned by $n^\mu$
and $\bar{n}^\mu$. 
In the framework of Soft-Collinear Effective Theory (SCET)~\cite{Bauer:2000ew,Bauer:2000yr,Bauer:2001ct,Bauer:2001yt}, the definition of the bare unpolarized quark fragmentation function~\cite{Collins:1981uw} takes on the following form~\cite{Procura:2009vm}: 
\begin{align} \label{eq:Dqdef}
  D_q^h(z) & = 
  \frac{1}{ z} \int\! \df^2 p_h^\perp \, \sum_X\, \frac{1}{2 N_c}\, \tr\,
  \frac{\bnslash}{2}  \de(p_{X h, r}^-)\, \de^2(p_{X h, r}^\perp)
  \nn \\
  & \times \Mae{0}{[\de_{\w, \bnP} \,\de_{0, \cP_\perp} \chi_n(0)]} {X h}
  \Mae{X h}{\bar \chi_n(0)}{0} \,,
\end{align}
in a frame where the parton $\perp$-momentum vanishes.
Here, $\chi_n$ is the $n$-collinear quark field that contains a Wilson line, making this definition (collinearly) gauge invariant. $N_c=3$ is the number of colors, and the trace is taken over color and Dirac indices.The $\bnP$ and $ {\cal{P}}_\perp$ operators~\cite{Bauer:2001ct} pick out the large, discrete label momentum of the field, while the continuous residual components of the jet momentum are denoted by $p_{X h, r}^\mu$. We use the notation $p_{X h}^\mu = p_X^\mu +p_h^\mu$. According to factorization at leading power, the sum
over the accompanying hadrons $X$ is dominated by jet-like configurations for the $|X h \rangle$ states~\cite{Collins:1989gx}.

The analysis in Ref.~\cite{Procura:2009vm}, which we shortly present here, combines fragmentation with the measurement of the invariant mass of the jet to which $h$ belongs. Since this probes fragmentation at a more differential
level, we expect it can teach us interesting things about the jet dynamics
involved in producing $h$.

%%%%%%%%%%%%%%%%%%%%%%%%%%%%%%%%%%%%%%
\section{the fragmenting jet function}

%There are three scales characterizing the process of interest 
We focus on processes governed by three different scales: the (perturbative) {\it hard scale} set by the jet energy $E_{X h}$, the intermediate (perturbative) {\it jet scale} given by the jet invariant mass $m_{X h}$, and the {\it soft scale}, of order $m_{X h}^2/E_{X h}$, with the hierarchy $m_h \ll m_{X h} \ll E_{X h}$.

Consider for example, without any loss of generality, a semi-leptonic $B$-decay in the endpoint region, where a single $u$-quark jet in the final state recoils against leptons. In the inclusive case, $\bar{B} \to X_u \ell \bar{\nu}$, the jet momentum $p_{X_u} ^\mu=(p_{X_u} ^+, p_{X_u} ^-, p_{X_u}^\perp)$, scales as $p_{X_u}^\mu \sim (\Lambda_{\rm QCD}, m_b, \sqrt{m_b\, \Lambda_{\rm
    QCD}})=m_b(\lambda^2, 1, \lambda) $ where $\lambda \sim \sqrt{\Lambda_{\rm
    QCD}/m_b}$ is the SCET expansion parameter. At leading order the following factorization theorem holds (see {\it e.g.}~\cite{Korchemsky:1994jb,Bauer:2001yt,Ligeti:2008ac}):
\begin{align} \label{eq:inclFT}
 \frac {d^2 \Gamma}{d p_{X_u}^+\, d p_{X_u}^-} 
  &= 2 (2 \pi)^3\, \Gamma_0\: H(m_B,p_{X_u}^-,p_{X_u}^+, \mu)\, p_{X_u}^- 
  \nn \\
  &\!\!\!\!\!\!\!\! \times \int_0^{p_{X_u}^+} dk^+
   J_u\big(k^+ p_{X_u}^-,\mu\big) S(p_{X_u}^+ - k^+,\mu) \,,
\end{align}
with the constant $\Gamma_0$ proportional to $| V_{ub}|^2$, the hard function $H$ for the underlying $b\to u\ell\bar\nu_\ell$ process, the inclusive jet function $J$ encoding the contribution from momenta which scale collinearly, and the non-perturbative shape
function $S$ which is the parton distribution for a $b$-quark in the $B$-meson
in the heavy quark limit. Here  $\mu$ denotes the $\overline{\rm MS}$-scheme renormalization scale. In Eq.(\ref{eq:inclFT}) the convolution variable $k^+$ is the plus-momentum of the up-quark initiating the $X_u$ jet.

Consider now the case of  $\bar{B} \to (X \pi)_u \ell \bar{\nu}$ where one light energetic hadron (a pion) is observed in the final state jet. In Ref.~\cite{Procura:2009vm} we proved the following leading-power factorization theorem:
%%%
\begin{align} \label{eq:semiinclFT}
\frac{d ^3 \Gamma}{d p_{X \pi}^+\, d p_{X \pi}^- \,d z} &= \Gamma_0\,
H(m_B, p_{X \pi}^-, p_{X \pi}^+,\mu )\, p_{X \pi}^- 
\nn \\
&\!\!\!\!\!\!\!\!\!\!\!\!\!\!\!\!\!\!\!\!\!\! \times \int_0^{p_{X \pi}^+} dk^+ \,
 {\cal G}_u^\pi\big(k^+p_{X \pi}^-,\,z, \mu \big)\,
 S(p_{X \pi}^+ - k^+,\mu)\,,
\end{align}
%%%
where $z$ is the fragmentation variable $p_\pi^-/p_{X \pi}^-$, and $\Gamma_0$, $H$ and $S$ are the same as in Eq.(\ref{eq:inclFT}). The energetic pion fragments from the jet and has a
collinear scaling $p_\pi^\mu\sim (\Lambda_{\rm QCD}^2/m_b, m_b, \Lambda_{\rm QCD})$: therefore $z$ counts as a quantity of order 1 and cannot be too small.  $\cG_{q}^h(s,z)$ is called a ``fragmenting jet function"~\cite{Procura:2009vm} since it represents an interpolating object between the jet function $J_q(s)$ and the standard fragmentation function $D_q^h(z)$ as we shall clarify: 
%%%
\begin{align} \label{eq:Gqdef}
  \cG_{q}^h(s,z) & = 
  \int\! \df^4 y\, e^{\img k^+ y^-/2}\, \int\! \df p_h^+
\, \sum_X \,\frac{1}{4N_c}\, 
 \nn\\ 
 & \times \tr \,
  \Big[\frac{\bnslash}{2} \Mae{0}{[\de_{\w,\bnP}\, \de_{0,\cP_\perp} \chi_n(y)]} {X h}
  \Mae{X h}{\bar \chi_n(0)}{0} \Big]
  \nn \\
 & = 
 \frac{(2\pi)^3}{ p_h^-} \int\! \frac{\df y^-}{2\pi}\, e^{\img k^+ y^-/2}
 \int\! \df^2 p_h^\perp\, \sum_X  \frac{1}{2 N_c}
 \nn\\ 
 & \times \tr  
 \Big[\frac{\bnslash}{2} \de(p_{X h, r}^-)\, \de^2(p_{X h, r}^\perp)
 \nn \\
 & \;\;\;\;\;\; \Mae{0}{[\de_{\w,\bnP}\, \de_{0,\cP_\perp} \chi_n(y^-)]} {X h}
 \Mae{X h}{\bar \chi_n(0)}{0} \Big]~.
\end{align}
%%% 
We work in a frame where $p_{X h}^\perp =0$. The integration over $y^-$ fixes the the partonic jet invariant mass $\sqrt{s}$, since $s=k^+ \w$. 

Our analysis yields the following simple replacement rule which allows to write factorization formulae for semi-inclusive processes from the corresponding inclusive ones:
%%%
\beq \label{repl}
J_i(s,\mu) \to \frac{1}{2 (2 \pi)^3}\, \cG_i^h(s,z,\mu)\, dz~.
\eeq
%%%
An example of this relation can be found by comparing Eq.(\ref{eq:inclFT}) with Eq.(\ref{eq:semiinclFT}). 
In Eq.(\ref{repl}) the factor $2 (2 \pi)^3$ is related to how we normalized $\cG_i^h$ and incorporates the phase space for the observed hadron $h$. This replacement rule also implies that the renormalization and the RG evolution of these two functions are the same. In particular, the renormalization of $\cG_i^h$ does not affect its $z$-dependence and does not mix quark and gluon fragmenting jet functions, at any order in perturbation theory.  $\cG_i^h$  does depend on two invariant mass scales, $s$ and $\Lambda_{\rm QCD}^2$, which we factorize below.

Furthermore, if we sum over all possible hadrons $h\in {\cal H}_i$ fragmenting from a parton $i$ and belonging to the jet, then the fragmenting jet function can be related to the inclusive jet function $J_i(s, \mu)$ which is completely calculable in perturbation theory.
One starts from the completeness relation
%%%
\beq \label{eq:Xh}
\int_0^1 \df z\,z\,\sum_{h \in {\cal H}_i} \sum_X  |X h(z)\rangle \langle X h(z)| = \sum_{X_i} |X_i \rangle \langle X_i|= \dblone\,,
\eeq
%%%
where the factor $z$ under the integral is explained in Ref.~\cite{JPW}, and is necessary to provide the correct symmetry factor for states with identical particles. $\{ |X_i \rangle \}$ is a complete set of states in the jet-like kinematic region that we are interested in. Applying Eq.(\ref{eq:Xh}) to the fragmentation function leads to
\begin{align} \label{eq:D_mom_cons}
  \sum_h \int_0^1 \df z\, z\, D_j^h(z,\mu) = 1
  \,,
\end{align}
which is consistent with momentum conservation and with the definition of $D_i^h(z,\mu)$ as the {\it number density} of the hadron $h$ in the parton $i$~\cite{Collins:1981uw}. In the case of the fragmenting jet function, Eq.(\ref{eq:Xh}) yields
%%%
\beq
  \sum_{h \in {\cal H}_i} \int_0^1 \df z\, z\; \cG_i^h(s,z,\mu) = 2(2\pi)^3 J_i(s,\mu)\,.
\eeq
%%%

$\cG^h (s,z,\mu)$ can be explicitly related also to $D^h(z, \mu)$. Indeed,
by performing an operator product expansion about the $y^- \to 0$ limit on the right-hand-side of Eq.(\ref{eq:Gqdef}), we can match onto the low-energy matrix elements that correspond to the fragmentation functions. This amounts to a \SCETa onto \SCETb matching at the intermediate scale set by the jet invariant mass $\sqrt{s}$:
%%%
\begin{align} \label{eq:OPE}
  \cG_i^h(s,z,\mu) 
  & = \!\!\sum_{j=g,u, \bar u, d, \dots} \int_z^1 \frac{\df z'}{z'}
   \cJ_{ij}\Big(s,\frac{z}{z'},\mu\Big) D_j^h(z',\mu)
   \nn \\
   & \times \bigg[1+ \ORd{\frac{\lqcd^2}{s}}\bigg]
  \,.
\end{align}
%%%
The perturbatively calculable Wilson coefficients $\cJ_{ij}$ are not sensitive to infrared physics, since they receive contributions from momenta of the order of the virtuality of the parton originating the jet. The dependence on $z$ and $z'$ is only through their ratio, since the $\cJ_{ij}$ can only depend on the perturbative variables associated with the partons $i$ and $j$, and not on the hadron $h$. The matching coefficients 
${\cal J}_{ij}$ describe the formation of a final state jet within which the non-perturbative, long-distance fragmentation process described by $D_j^h$ takes place.

The one-loop calculation of ${\cal J}_{ij}$ for both quark and gluon fragmentation is the subject of Ref.~\cite{JPW}. This calculation completes the picture detailed in Ref.~\cite{Procura:2009vm} with the information necessary to relate the factorization theorems for semi-inclusive processes where the jet invariant mass is probed with the standard $D_i^h(z,\mu)$ up to NLO/NNLL accuracy. As an example,  by applying Eq.~(\ref{eq:OPE}) to the factorization theorem for $e^+e^-\to \mbox{(dijets)}$~\cite{Korchemsky:1999kt,Korchemsky:2000kp,Fleming:2007qr,Schwartz:2007ib} we obtain the following factorized differential cross-section for $e^+e^- \to \mbox{(dijets)}+h$~\cite{Procura:2009vm}:
%%%
\begin{align}
  &\frac{d^3 \sigma}{d M^2\, d \bar{M}^2\, d z}
   = \frac{\sigma_0}{2(2 \pi)^3}H_{\rm 2jet}(Q, \mu)  \sum_j \int_{-
    \infty}^{+ \infty}\! d l^+\, d l^- 
  \nn \\
  &\;\;\;\;\;\;\;\;  \int_z^1 \frac{dx}{x}\, \Big[ 
  \!{\cal J}_{q j}\!\Big(M^2 - Q l^+, \frac{z}{x}, \mu\Big) \,
   J_{\bar{n}}\! \big(\bar{M}^2 -Q l^-, \mu \big)
   \nn \\
  &\;\;\;\;\;\;\;\; +
 J_{n}\! \big({M}^2 -Q l^+, \mu \big)
 {\cal J}_{\bar q j}\!\Big(\bar M^2 - Q l^-, \frac{z}{x}, \mu\Big) 
 \! \Big]  
 \nn \\
  &\;\;\;\;\;\;\;\;  \times D_j^h(x, \mu) S_{\rm 2jet}(l^+, l^-, \mu)\, ,
\end{align}
%%%
where $\sigma_0$ is the tree level total cross-section which acts as a
normalization factor, $Q$ is the center-of-mass energy, $M^2$ and $\bar M^2$ are
hemisphere invariant masses for the two hemispheres perpendicular to the dijet
thrust axis. Since here we assume that it is not known whether the hadron $h$
fragmented from the quark- or antiquark-initiated jet, we have a sum over both
possibilities in the factorization theorem. For the definitions of $\sigma_0$,
$H_{\rm 2jet}$, and $S_{\rm 2jet}$ see Ref.~\cite{Fleming:2007qr} whose
notation we have followed.
We refer the reader to \cite{JPW} for an analysis to NNLL accuracy of the process $e^+ e^- \to X \pi^+$ on the $\Upsilon(4S)$ resonance where one measures the momentum fraction of the $\pi^+$ and restricts to the dijet limit by imposing a cut on thrust. 

The factorization formulae derived with our analysis should enable improved
constraints on parton fragmentation functions to light hadrons, by allowing
improved control over the fragmentation environment with the invariant mass
measurement, as well as opening up avenues for fragmentation functions to be
measured in new processes, such as $B$-decays, see~\cite{Procura:2009vm}.  We also expect that further
study based on the definition of the fragmenting jet functions, will contribute to a better
understanding of the relative roles of perturbative partonic short-distance
effects and non-perturbative hadronization in shaping jet
features.

\begin{theacknowledgments}
We acknowledge support by the Office of Nuclear Physics of the U.S.\ 
Department of Energy under the Contract DE-FG02-94ER40818, and by the Alexander
von Humboldt foundation through a Feodor Lynen Fellowship (M.P.) and a Friedrich
Wilhelm Bessel award (I.S.).
\end{theacknowledgments}

\end{document}